\documentclass{emulateapj}
\begin{document}
\title{ANALYSIS OF SPATIAL STRUCTURE OF THE SPICA \ion{H}{2} REGION}
\author{J.-W. PARK\altaffilmark{1,2}} \author{K.-W.
MIN\altaffilmark{1}} \author{K.-I. SEON\altaffilmark{3}}
\author{W. HAN\altaffilmark{3}} \author{J.
EDELSTEIN\altaffilmark{4}}

\altaffiltext{1}{Korea Advanced Institute of Science and Technology
(KAIST), 305-701 Daejeon, Republic of Korea; jwp@kari.re.kr}

\altaffiltext{2}{Korea Aerospace Research Institute (KARI), 305-333
Daejeon, Republic of Korea}

\altaffiltext{3}{Korea Astronomy and Space Science Institute (KASI),
305-348 Daejeon, Republic of Korea}

\altaffiltext{4}{Space Sciences Laboratory, University of
California, Berkeley, CA 94720, USA}

\begin{abstract}
Far ultraviolet (FUV) spectral images of the Spica \ion{H}{2} region
are first presented here for the \ion{Si}{2}$^{\ast}$
$\lambda$1533.4 and \ion{Al}{2} $\lambda$1670.8 lines and then
compared with the optical H$\alpha$ image. The H$\alpha$ and
\ion{Si}{2}$^{\ast}$ images show enhanced emissions in the southern
part of the \ion{H}{2} region where \ion{H}{1} density increases
outwards. This high density region, which we identify as part of the
``interaction ring" of the Loop I superbubble and the Local Bubble,
seems to bound the southern \ion{H}{2} region. On the other hand,
the observed profile of \ion{Al}{2} shows a broad central peak,
without much difference between the northern and southern parts,
which we suspect results from multiple resonant scattering. The
extended tails seen in the radial profiles of the FUV intensities
suggest that the nebula may be embedded in a warm ionized gas.
Simulation with a spectral synthesis code yields the values of the
Lyman continuum luminosity and the effective temperature of the
central star similar to previous estimates with $10^{46.2}$ photons
s$^{-1}$ and 26,000 K, respectively, but the density of the northern
\ion{H}{2} region, 0.22 cm$^{-3}$, is much smaller than previous
estimates for the H$\alpha$ brightest region.
\end{abstract}

\keywords{ISM: individual (Spica)  --- \ion{H}{2} region
--- ultraviolet: ISM}

\section{INTRODUCTION}
\ion{H}{2} regions are generally located in the vicinity of OB stars
because these hot stars can produce strong ultraviolet radiation,
thereby photoionizing hydrogen atoms in the region. The energy
balance between photoelectrons and forbidden line cooling sets the
gas temperature to $\sim10^{3.8}$ K in the \ion{H}{2} region
\citep{ost89}. The structure of \ion{H}{2} region has been studied
in terms of the Balmer recombination lines of hydrogen atoms, the
observations of other optical forbidden lines (e.g., [\ion{S}{2}]
$\lambda$6716, [\ion{O}{2}] $\lambda$3727)
\citep[e.g.,][]{pei93,pei03,wan04} and infra-red fine-structure
lines (e.g., [\ion{O}{3}] 88 $\mu$m, [\ion{S}{3}] 33 $\mu$m)
\citep[e.g.,][]{mar02}. The theoretical advances have also been made
by means of various photoionization models
\citep[e.g.,][]{sta82,erc03,mor05}. Our understanding of the
\ion{H}{2} region has been advanced considerably in recent times
with the aid of the Wisconsin H$\alpha$ Mapper (WHAM) data
\citep{haf03}. With a velocity resolution capability of 8$-$12 km/s,
WHAM measured the H$\alpha$ emission from warm ionized objects
within $\sim\pm$100 km s$^{-1}$ of the Local Standard of Rest and
provided the first large-scale H$\alpha$ survey, covering the
three-quarters of the northern sky. The survey results show that
enhancements can generally be seen near the planetary nebulae and
\ion{H}{2} regions surrounding massive O and early B-type stars ;
the results also confirm the presence of unidentified high galactic
components \citep{rey05}.

The $\alpha$ Vir (Spica), one of the brightest stars at a high
galactic latitude, was found to be a double-lined spectroscopic
binary with spectral types B1 V and B4 V \citep{her71}. The
existence of the \ion{H}{2} region around $\alpha$ Vir was suggested
because the region appeared to be a hole in radio observations
\citep{fej74} and ultraviolet absorption lines were seen for the
star of the nearby sightline \citep{yor79}. Equipped with a
Fabry-Perot spectrometer, \citet{rey85} made H$\alpha$ scans and
revealed that the region was indeed ionized with a gas density of
$\sim$0.6 cm$^{-3}$. The hydrogen ionization rate of the whole
\ion{H}{2} region was estimated to be $\sim$ $10^{46.3}$ photons
s$^{-1}$. The ratio of [\ion{S}{2}] $\lambda$6716 to the H$\alpha$
line, which signifies the contribution of collisional excitation
with enhanced temperature, was also made for the Spica \ion{H}{2}
region (Spica Nebula). While the Spica Nebula is generally accepted
as a normal \ion{H}{2} region, the ratio was found to be rather high
(0.16 at the center and 0.21 at the edge), compared to those of
other \ion{H}{2} regions such as the Orion Nebula
\citep[0.019,][]{pei77} and the Sharpless 261
\citep[0.059,][]{haw78}, though its origin was not clearly
identified \citep{rey88}. Recently, the Spica Nebula was observed in
the WHAM survey \citep{rey04}, but no detailed study of this set of
data has yet been published.

\ion{Si}{4} and \ion{C}{4} ion lines were detected in the far
ultraviolet (FUV) absorption line study towards Spica, and their
origin was ascribed to the Local Bubble (LB) surrounding the Sun
\citep{sav09}. Being a photoionized \ion{H}{2} region, the Spica
Nebula may not be associated with these high-stage FUV lines, but it
can still be a source of FUV emission from low-stage ions,
especially in view of the high [\ion{S}{2}] $\lambda$6716 to
H$\alpha$ line ratio. In this paper, we analyze the FUV
\ion{Si}{2}$^{\ast}$ and \ion{Al}{2} emission lines as well as the
WHAM survey data, to study the \ion{H}{2} region around Spica. It
should be noted that the ionization potentials of Si$^{2+}$ and
Al$^{2+}$ are similar to that of hydrogen: 16.4 eV for Si$^{2+}$ and
18.8 eV for Al$^{2+}$. As the spatial structure of the Spica
\ion{H}{2} region was not explored previously and there have been no
reports of the emission line study in the FUV wavelengths, we
believe the present study should provide useful information about
the global morphology of the nebula. For the FUV study, we use the
same data set as the one used for our previous analysis of Loop I
superbubble \citep[L1,][]{park07} ; that data set was obtained from
the Far-ultraviolet Imaging Spectrograph (FIMS) on the Korean
microsatellite \textsl{STSAT-1} \citep{ede06}. The FIMS is an
instrument optimized for the measurement of diffuse FUV emissions
with a large field of view ($7\fdg5\times4\farcm3$) for the
wavelength band of 1330$-$1720{\AA}. We also compare the
observational results with the results of the Cloudy photoionization
model so that we can constrain the physical parameters associated
with the \ion{H}{2} region and $\alpha$ Vir itself.

\section{OBSERVATIONS}
In Figure 1, we plot spectra for the three distinct regions of the
Spica Nebula : the core region within $0\fdg5$ circle including the
central point source, $\alpha$ Vir, in the top panel ; the nebula
region defined by $0\fdg5-8\fdg0$ circle \citep{rey85} in the middle
panel; and the background of a 12$\arcdeg\times$12\arcdeg square
outside the nebula in the bottom panel. The spectra were binned with
1 {\AA} and smoothed with a 3 {\AA} wide boxcar. For comparison, we
also include in the top panel the spectrum for $\alpha$ Vir as
observed by International Ultraviolet Explorer (\textsl{IUE}).

As can be seen in the top panel, the prominent absorption lines of
\ion{Si}{4} $\lambda\lambda$ 1393.8, 1402.8 in the \textsl{IUE}
spectrum, which undoubtedly originate from the stellar atmosphere,
appear as strong emission lines in the diffuse FIMS spectrum ; these
lines indicate the extended hot gas around the central star. Similar
features can also be identified for \ion{Si}{2}$^{\ast}$
$\lambda$1533.4 and the \ion{C}{4} doublet $\lambda\lambda$ 1548.2,
1550.8. However, as shown in the middle panel, the \ion{Si}{4}
emission lines are not seen in the nebula region while
\ion{Si}{2}$^{\ast}$, \ion{C}{4} doublet and \ion{Al}{2}
$\lambda$1670.8 are bright. The \ion{C}{4} doublet is also
conspicuous in the background region (bottom panel) while both the
\ion{Si}{2}$^{\ast}$ and \ion{Al}{2} lines become less prominent
compared to the nebula region. Hence, only the \ion{Si}{2}$^{\ast}$
and \ion{Al}{2} may actually be the dominant emission lines in the
nebula region as the \ion{C}{4} doublet seen in the middle panel
could be contributed from the projected background.

We have constructed FUV spectral images of the Spica Nebula
\citep[radius of $\sim$8\arcdeg,][]{rey85}, extended to
12\arcdeg$\times$12\arcdeg to include the nearby background medium.
The \ion{Si}{2}$^{\ast}$ and \ion{Al}{2} images were made by
utilizing the HEALPix scheme \citep{gor05} with a pixel resolution
of $\sim0\fdg92$. The \ion{Si}{2}$^{\ast}$ and \ion{Al}{2} lines
were fitted with single Gaussian profiles in the spectral range
1520$-$1546 {\AA} and 1658$-$1684 {\AA}, respectively, for each
pixel. The images were smoothed with a Gaussian function whose full
width at half-maximum (FWHM) was 3\arcdeg. The resulting signal to
noise (S/N) ratios for the bright features of the nebula are above
3.0 for both \ion{Si}{2}$^{\ast}$ and \ion{Al}{2}.

Figure 2 shows the final \ion{Si}{2}$^{\ast}$ and \ion{Al}{2} images
taken from FIMS. It also displays the H$\alpha$ image taken from
\citet{fin03} and the \ion{H}{1} map taken from \citet{ka05}. We
designated $\alpha$ Vir in these images with an asterisk at
$(RA,DEC) = (201\fdg3, -11\fdg2)$. The black circles with a radius
of 8\arcdeg, correspond to 12 pc for a distance of 80 pc to $\alpha$
Vir; they indicate the region conventionally defined as the Spica
Nebula \citep{rey85}. First, we note that the \ion{Si}{2}$^{\ast}$
image in Figure 2a shows an asymmetric feature with strong
enhancement in the southern region below $\alpha$ Vir. The southern
enhancement is also seen in the H$\alpha$ map of Figure 2c and seems
to be related to the high neutral hydrogen density shown in Figure
2d. This asymmetric feature is less clear in the \ion{Al}{2} map of
Figure 2b; it does not show much difference between the northern and
southern parts though the image seems to extend from the
round-shaped central peak to the northwest direction, where the
emission of H$\alpha$ is somewhat enhanced. In Figure 2c, we
overplotted the \ion{H}{1} contours from $2.0\times10^{20}$
cm$^{-2}$ to $8.0\times10^{20}$ cm$^{-2}$ with $2.0\times10^{20}$
cm$^{-2}$ intervals: $<$ $2.0\times10^{20}$ cm$^{-2}$ generally
corresponds to the northern region above $\alpha$ Vir;
$2.0\times10^{20}$ $\sim$ $6.0\times10^{20}$ cm$^{-2}$ represents a
band that passes through the southern \ion{H}{2} region; and $>$
$6.0\times10^{20}$ cm$^{-2}$ represents the area below the Spica
Nebula. In Figure 2d, a \ion{H}{1} cavity is seen in the vicinity of
$\alpha$ Vir which is probably generated by the strong Lyman
continuum from the star that causes almost all of the neutral
hydrogen atoms to be ionized.

We constructed radial profiles of the H$\alpha$,
\ion{Si}{2}$^{\ast}$ and \ion{Al}{2} intensities for the Spica
Nebula. Because the H$\alpha$ and \ion{Si}{2}$^{\ast}$ images show
strong asymmetry, we made the northern and southern intensity
profiles separately by averaging the intensity for the corresponding
concentric semicircles of 1\arcdeg bins up to 8\arcdeg. We
subtracted the background radiation in these profiles: 1.4 Rayleigh
(R) for H$\alpha$, the average value of the galactic latitudes
between 40$\arcdeg$ and 60$\arcdeg$ \citep{rey84}; 0.03 R for
\ion{Si}{2}$^{\ast}$ and 0.04 R for \ion{Al}{2} where 1R = 8.0
$\times$ 10$^{4}$ photons cm$^{-2}$ s$^{-1}$sr$^{-1}$. We also made
extinction corrections by using the extinction curve given by
\citet{car89} though the corrections are not significant. With an
assumed \textsl{E(B-V)} value of 0.01 for the \ion{H}{2} region, the
extinction corrected intensities of H$\alpha$, \ion{Si}{2}$^{\ast}$
and \ion{Al}{2} are respectively, about 1.02, 1.08 and 1.07 times
the observed intensities \citep{gal08}.

The top panel in Figure 3 shows the resulting profiles. The solid
line represents the northern profiles and dashed lines represents
the southern profiles. The rather large error bars come from the
significant spatial variation that still exists within each
semicircle bin. As expected, in the H$\alpha$ and
\ion{Si}{2}$^{\ast}$ intensities we see a marked difference between
the northern and southern regions with a much slower decrease for
the southern profiles; in contrast, the \ion{Al}{2} intensity does
not show a marked difference between the northern and southern
profiles. The southern H$\alpha$ profile even shows a flat region
from 3 pc to 8 pc, which is undoubtedly due to the fact that the
density increases with distance. The bottom panel in Figure 3 shows
the model profiles obtained from a photoionization simulation and
will be discussed in later section.

In Figure 4 we show a scatter plot of \textit{N}(\ion{H}{1}) against
the H$\alpha$ intensity. A correlation is seen for
\textit{N}(\ion{H}{1}) below $6.0\times10^{20}$ cm$^{-2}$ and the
0.62 correlation coefficient confirms that the H$\alpha$ enhancement
in the southern Spica region can be attributed to that region's high
neutral hydrogen density. Hence, the Spica Nebula is
ionization-bounded in the southern region with an ionization front
at \textit{N}(\ion{H}{1}) = $6.0\times10^{20}$ cm$^{-2}$. The
H$\alpha$ intensity drops significantly to 1.6 R (Log H$\alpha$=0.2)
for \textit{N}(\ion{H}{1}) $>$ $6.0\times10^{20}$ cm$^{-2}$ (Log
\textit{N}(\ion{H}{1})=20.78); and the intensity drop implies that
any H ionization photons that pass through this dense ambient medium
are dissipated in the medium. It should be noted that the 1.6 R
value that we obtained here is very similar to the average value
(1.4 R) for the galactic latitudes between 40$\arcdeg$ and
60$\arcdeg$, which were estimated by \citet{rey84}.

\section{DISCUSSION}
The observed intensity profiles of the northern and southern
\ion{H}{2} regions shown in the top panel of Figure 3 are compared
with model calculations for the corresponding regions. When we ran
the photoionization simulation code Cloudy, we varied the Lyman
continuum luminosity, Q(\textsl{H}), the effective temperature of
the central star, and the density of the ambient medium. The
simulation was performed for a spherically symmetric nebula by using
the stellar atmospheric model of \citet{cas04} and by assuming the B
star abundance reported by \citet{kil94} and \citet{sem00}. We
obtained the volume emissivity as a function of the radius and
calculated the intensity toward a line of sight defined by the
offset angle from the central star. Finally, we smoothed the
profiles with a Gaussian function whose FWHM was 3\arcdeg for a
direct comparison with the observations.

First, the calculation was performed for a constant density medium,
corresponding to the northern region where density is more or less
uniform. The results are shown as solid lines in the bottom panel of
Figure 3. The model that fits best the northern H$\alpha$ profile
was obtained with Q(\textsl{H}) = $10^{46.2}$ photons s$^{-1}$,
effective temperature of 26,000 K, and the density of 0.22
cm$^{-3}$. These values of Q(\textsl{H}) and the effective
temperature are quite similar to previous estimates of $10^{46.3}$
photons s$^{-1}$ in \citet{rey85} and 25,791 K in \citet{kun97},
respectively. However, the density is much smaller than the result
of 0.6 cm$^{-3}$ estimated by \citet{rey85}, probably because the
observation was made for the brightest portion of the nebula which
is located in the central region. On the other hand, the model
profiles of \ion{Si}{2}$^{\ast}$ and \ion{Al}{2} are quite different
from the corresponding observed profiles, especially in the outer
region beyond 3 pc, where the observed profiles show a much slower
decrease with distance and have extended tails. In addition, the
\ion{Al}{2} model profile shows a much higher central peak intensity
with a sharper decrease than the observed profile even inside 3 pc.
These aspects will be discussed later, together with the results of
the southern profiles.

Next, we modeled the southern region in which \textit{N}(\ion{H}{1})
is seen to increase outwards. As our goal was to study the general
trends affected by the density gradient, not to reproduce the
observed profiles exactly, we used the power-law model built in the
code Cloudy, even though it may not represent the density profile
accurately:
\begin{equation}
n(r) = n_{c}(\frac{r}{r_{0}})^{-\alpha},
\end{equation}
where $n_{c}$ is the density at $r_{0}$, which was taken to be 0.3
pc. Best fit was obtained by matching the H$\alpha$ profiles with
fixed Q(\textsl{H}) and effective temperature from the simulation
for the northern region. The resulting value is $\alpha$ $=$ -0.15
(corresponding to the density of 0.38 cm$^{-3}$ at r = 12 pc), and
the model profiles are shown as dashed lines in the bottom panel of
Figure 3. The H$\alpha$ fit looks more or less reasonable although
the model shows a little higher central peak with a sharper
decrease, which probably originates from the limits imposed by the
power law profile of the model. However, the most significant
discrepancies seem to be in the \ion{Si}{2}$^{\ast}$ and \ion{Al}{2}
profiles. For example, the model \ion{Si}{2}$^{\ast}$ profile does
not show such an enhancement as the one seen in the observed profile
region between 2 pc and 8 pc, not responding to the outward density
increase. The density gradient effect is not manifested in the model
\ion{Al}{2} result, either, with a profile very similar to that of
the northern region. In fact, for \ion{Al}{2}, both the northern and
southern observations look similar, with smaller central peaks and
broader profiles than the models.

Hence, several discrepancies exist between the observation results
and those of a simple photoionization model such as Cloudy. For
example, the observed \ion{Si}{2}$^{\ast}$ and \ion{Al}{2} profiles
show extended tails not reproduced by Cloudy. The observed
\ion{Si}{2}$^{\ast}$ profile shows the effect of the increasing
density gradient in the southern region of the nebula while the
model does not reproduce the feature. The observed \ion{Al}{2}
profile, compared to the model profile, is significantly broader
with a much smaller central peak (about 25\% of the model). In
addition, the observed \ion{Al}{2} profile does not respond to the
density gradient. We would like to discuss these peculiar features
here. First, we believe the extended tails seen in the observed
profiles are caused by the background of enhanced temperature.
Figure 1 clearly shows the existence of the \ion{C}{4} doublet in
the nebula regions as well as in the background, which can hardly be
produced by photoionization in \ion{H}{2} regions. It should also be
noted that \ion{Si}{4} and \ion{C}{4} ions were detected recently
towards Spica \citep{sav09}. We will argue later that the Spica
Nebula is located close to the interface between the L1 and LB,
which contain hot gases. In addition to this background hot gas, the
Spica Nebula may actually be embedded in a medium of enhanced
temperature. The temperature profile obtained from our
photoionization model for the southern region of the nebula shows a
temperature drop from $\sim$5,000 K to $\sim$3,000 K with a change
in distance from 1 pc to 4 pc, while it remains more or less at
$\sim$5,000 K in the northern region. In this regard, it should be
noted that the ratio of [\ion{S}{2}] $\lambda$6716 to the H$\alpha$
line was observed to be rather high in the Spica Nebula region with
its value of 0.21 at $(RA,DEC) = (205\fdg3, -11\fdg1)$
\citep{rey88}, as mentioned previously. Assuming the general line
ratio of 1.3 for [\ion{S}{2}] $\lambda$6716/$\lambda$6731
\citep{cha83}, the [\ion{S}{2}]$\lambda\lambda$6716 + 6731/H$\alpha$
is estimated to be 0.36. This value is rather high for a
photoionized \ion{H}{2} region and actually resides between those of
a general \ion{H}{2} region and a mixing layer of shocked gases
\citep{las77}. We suspect that the model \ion{Si}{2}$^{\ast}$
profile would have been more responsive to the density increase if
the ambient medium were assumed to be of enhanced temperature, the
situation which the present Cloudy model does not simulate.

The enhanced temperature may also cause broadening of the observed
\ion{Al}{2} profile. However, the profile, both in the observation
and in the model, does not show north-south asymmetry. Moreover, the
observed central peak is much smaller than that expected from the
model. We believe this discrepancy originates from the multiple
resonant scattering of \ion{Al}{2} in the nebula, which is not
properly considered in the photoionization model Cloudy. The
resonance lines, such as \ion{Al}{2}, undergo many resonant
scattering events, resulting in a much longer random walk before
they can escape the nebula. The numerous scatterings can yield a
consequently higher probability of being absorbed by dust. While
detailed calculation requires extensive scattering simulations which
are not pursued here, a simple estimation would demonstrate
reduction of \ion{Al}{2} intensity to the observed value. Using the
Al$^{+}$ column density 13.05 cm$^{-2}$ and the average nebular
temperature 10$^{3.6}$ K obtained from the current photoionization
model, the optical depth of the resonance scattering of \ion{Al}{2}
is estimated to be 22.9 \citep{ver06}. The number of scattering
within a medium would be approximately given by $\rm
N_{scatt}\thickapprox max(\tau,\tau^{2})$ \citep{ryb79}. Then, the
total path length of the random walking photons would be $\rm
L_{tot}\thickapprox \rm N_{scatt}\times \rm L/\tau=\rm L\times
max(1,\tau)$, where $\rm L$ is the typical nebular size, and the
resonant lines would experience $\tau$ times higher dust-extinction
than that of the original nebular size ($\tau > 1$). Adopting
\textsl{E(B-V)} $=$ 0.01 \citep{gal08}, the dust-extinction optical
depth of \ion{Al}{2} is 1.47, and thereby only $\sim$ 23\% of
\ion{Al}{2} line would escape from the nebula, and the value is
roughly consistent with the observed peak intensity. Another effect
of this multiple scattering is of course broadening of the radial
profile as the photons diffuse out radially through random walks.
The effect of increasing density in the southern region may also
have become obscured due to this diffusion.

We further note that the ground state of the singly ionized ion
Si$^{+}$ is split. The emission lines from the excited state
3$s^{2}$4$s^{1}$ $^{2}$S$_{1/2}$ to the true ground state
3$s^{2}$3$p^{1}$ $^{2}$P$_{1/2}$ (\ion{Si}{2}, 1527{\AA}) and to the
excited ground state 3$s^{2}$3$p^{1}$ $^{2}$P$_{3/2}$
(\ion{Si}{2}$^{\ast}$, 1533{\AA}) has a transition ratio of 1 : 2.
The true \ion{Si}{2} $\lambda$1527 ground state transition is
optically thick and therefore undetectable \citep{kor06}. Adopting
the same way for the \ion{Al}{2}, the optical depth of the resonance
line \ion{Si}{2} is estimated to be 12.6. However, resonantly
absorbed \ion{Si}{2} make downward transitions with the line ratio
of 1 : 2 between \ion{Si}{2} and \ion{Si}{2}$^{\ast}$, most of
\ion{Si}{2} are converted to \ion{Si}{2}$^{\ast}$ after only a few
resonance scatterings, which halts further resonant scattering ($\rm
N_{scatt} = 2; I_{1527}/I_{1533} \approx 0.04$). Therefore
\ion{Si}{2}$^{\ast}$ would not be significantly reduced by dust
absorption.

As mentioned above, the southern enhancement in the Spica Nebula is
clearly associated with the dense \ion{H}{1} region shown in Figure
2d, in which \textit{N}(\ion{H}{1}) increases from less than
$2\times10^{20}$ to over $7\times10^{20}$ cm$^{-2}$. This \ion{H}{1}
shell bounding the ionization front was initially suggested to be a
\ion{H}{1} cloud by \citet{fej74}. However, we would like to note
that this \ion{H}{1} shell coincides with the region of interaction
between the L1 and LB, called ``interaction ring" (IR)
\citep{egg95}. The estimated distance to the IR ($\sim$ 70 pc) is
comparable to that of the Spica Nebula \citep[68$-$92 pc,][]{rey85}.
Extended morphology of the IR near the Spica was recently discussed
in \citet{park07}. The result indicates that the warm ionized gas in
the Spica Nebula was possibly created in the already disturbed
medium by the IR which affects the evolution of the H$\alpha$ region
significantly.

\section{CONCLUSION}
The H$\alpha$ distribution as well as those of \ion{Si}{2}$^{\ast}$
and \ion{Al}{2} obtained in the FUV wavelengths was analyzed for a
12\arcdeg$\times$12\arcdeg sky around the Spica Nebula. The spatial
variation, shown in both H$\alpha$ and \ion{Si}{2}$^{\ast}$ images,
confirmed that the Spica Nebula is bounded by a high density
\ion{H}{1} shell in the southern region which we relate to the
``interaction ring" of the L1 and the LB. We have also argued that
the nebula is possibly embedded in a gas of enhanced temperature,
based on that the observed emission profiles have long extended
tails and they respond to the density increase in the southern
region of the nebula, which is not explained by photoionization
only.

The \textsl{IUE} data presented in this paper were obtained from the
Multimission Archive at the Space Telescope Science Institute
(MAST). STScI is operated by the Association of Universities for
Research in Astronomy, Inc., under NASA contract NAS5-26555. Support
for MAST for non-HST data is provided by the NASA Office of Space
Science via grant NAG5-7584 and by other grants and contracts.

{}
\clearpage

\begin{figure} \epsscale{0.7} \plotone{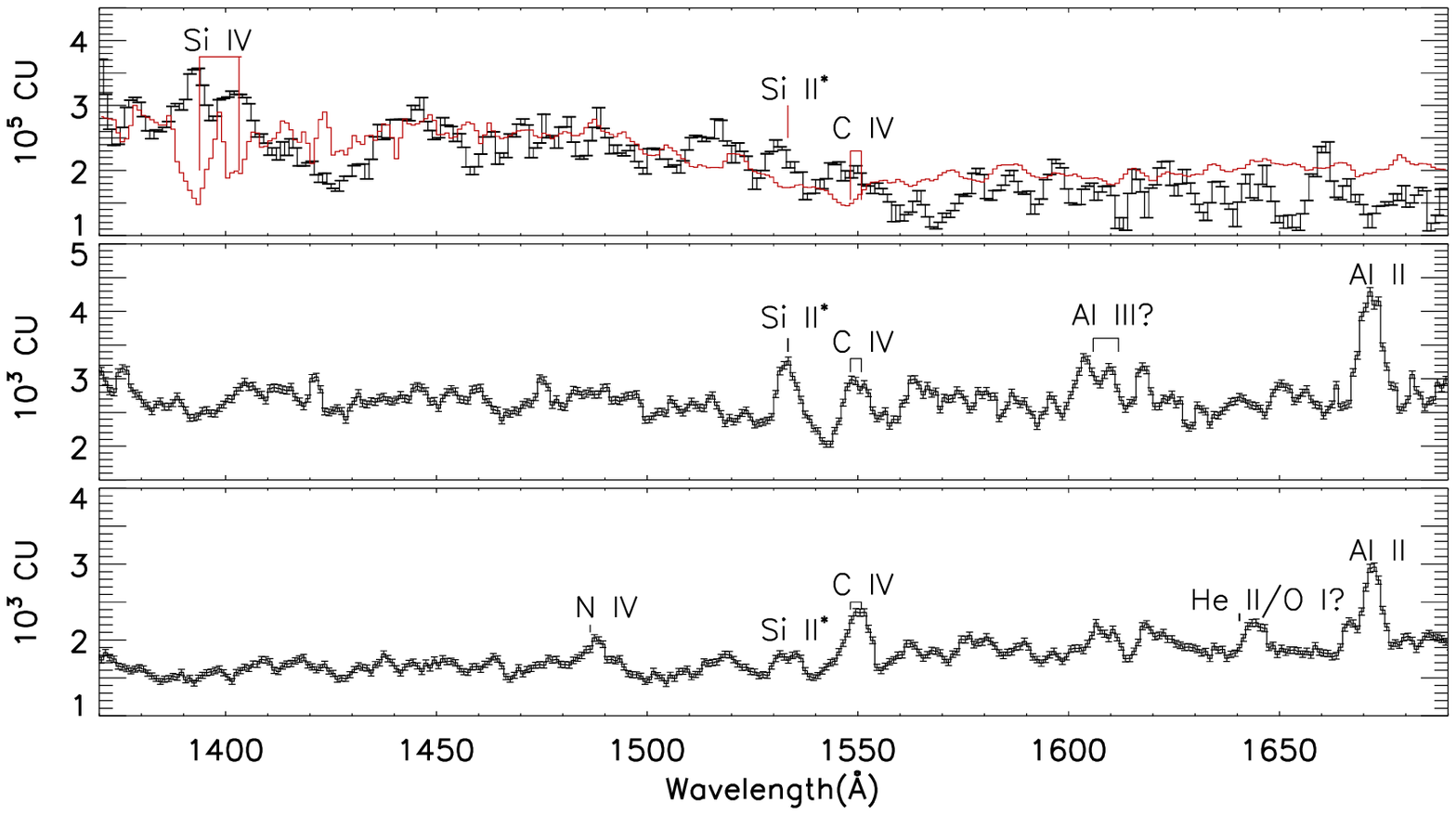}
\caption{FUV Spectra of the Spica Nebula: the core region within
$0\fdg5$ (top panel), the nebula region of $0\fdg5-8\fdg0$ (middle
panel), and the background of 12$\arcdeg\times$12\arcdeg square
outside the nebula (bottom panel). The spectrum of the central star
observed by \textsl{IUE} is also plotted in red in the top panel.
Several ion lines are marked in the figure. The spectra were binned
with 1 {\AA} and smoothed with a boxcar whose width is 3 {\AA}.
NOTE.- CU = photons s$^{-1}$ cm$^{-2}$ sr$^{-1}$ {\AA}$^{-1}$
\label{fig1}}
\end{figure}

\begin{figure} \epsscale{0.7} \plotone{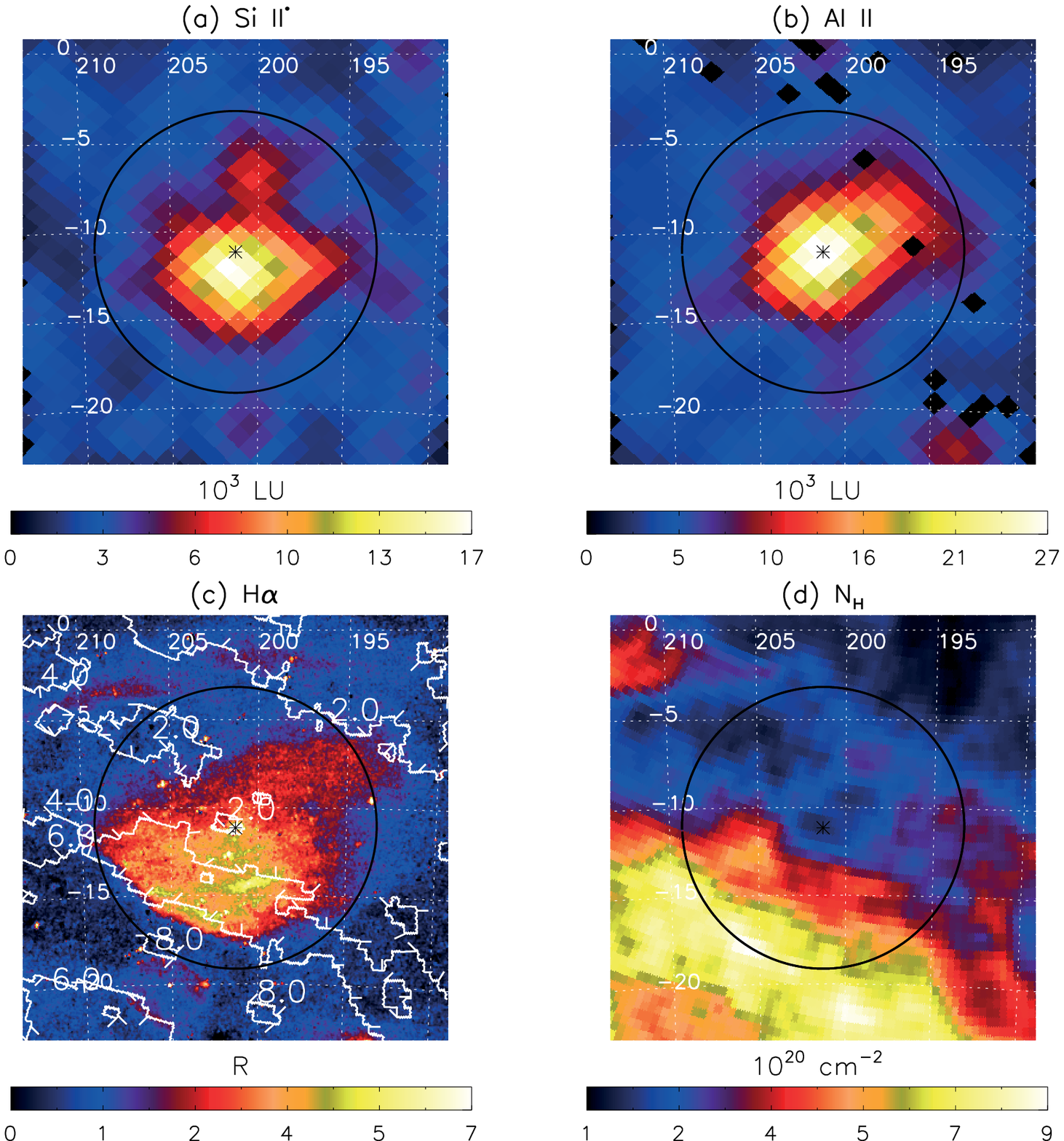}
\caption{Spectral images of the Spica \ion{H}{2} region: (a) FIMS
\ion{Si}{2}$^{\ast}$; (b) FIMS \ion{Al}{2}; (c) WHAM H$\alpha$
\citep{fin03}; and (d) $\textit{N}_{H}$ map \citep{ka05}. The large
black circles of radius of 8\arcdeg in (a) - (d) indicate the region
conventionally defined as the Spica Nebula. The $\textit{N}_{H}$
contours are overlaid in (c). NOTE.- LU = photons s$^{-1}$ cm$^{-2}$
sr$^{-1}$}
\end{figure}

\begin{figure} \epsscale{0.7} \plotone{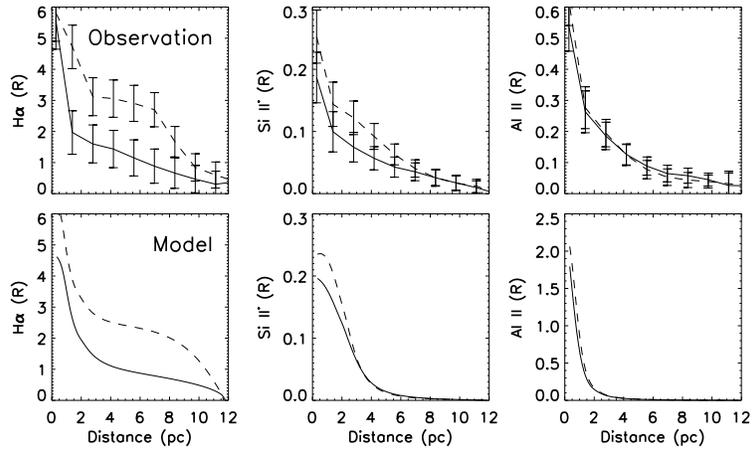}
\caption{Radial profiles of the H$\alpha$, \ion{Al}{2} and
\ion{Si}{2}$^{\ast}$ for the Spica Nebula. Top panel shows the
observed profiles with 1 sigma error bars for the northern (solid
lines) and southern (dashed lines) regions. The results of the
spectral synthesis models for the corresponding northern and
southern regions are shown in the bottom panel.}
\end{figure}

\begin{figure} \epsscale{0.6} \plotone{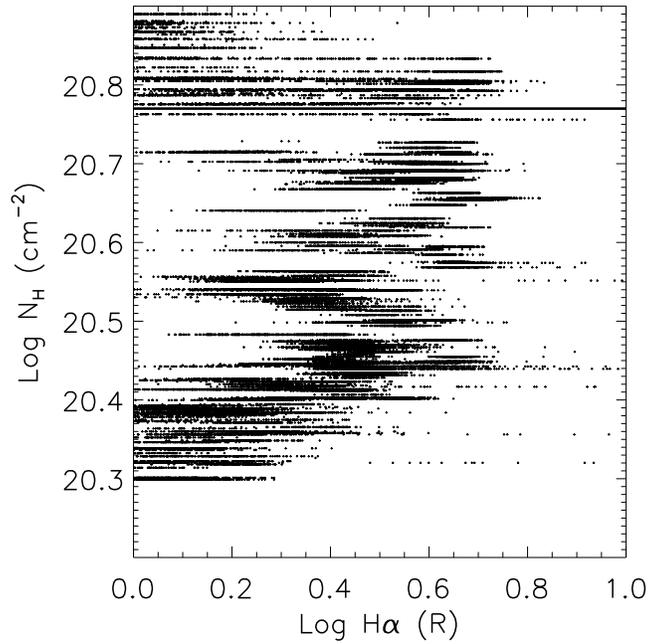}
\caption{\textit{N}(\ion{H}{1}) is plotted against H$\alpha$ in the
logarithmic scale for the Spica Nebula region 8\arcdeg. Correlation
is seen for \textit{N}(\ion{H}{1}) below $6.0\times10^{20}$
cm$^{-2}$ (marked as a black horizontal line). These data points
represent the correlation coefficient of 0.62.}
\end{figure}


\begin{thebibliography}{}
\bibitem[Cardelli et al.(1989)]{car89} Cardelli, J.~A.,
Clayton, G.~C., \& Mathis, J.~S.\ 1989, \apj, 345, 245
\bibitem[Castelli \& Kurucz(2004)]{cas04} Castelli, F., \& Kurucz,
R.~L.\ 2004, arXiv:astro-ph/0405087
\bibitem[Chanot \& Sivan(1983)]{cha83} Chanot, A., \& Sivan, J.~P.\
1983, \aap, 121, 19
\bibitem[Edelstein et al.(2006a)]{ede06} Edelstein, J., et al. 2006a, \apj, 644, L153
\bibitem[Egger \& Aschenbach(1995)]{egg95} Egger, R.~J., \&
Aschenbach, B.\ 1995, \aap, 294, L25
\bibitem[Ercolano et al.(2003)]{erc03} Ercolano, B., Barlow,
M.~J., Storey, P.~J., \& Liu, X.-W.\ 2003, \mnras, 340, 1136
\bibitem[Fejes(1974)]{fej74} Fejes, I.\ 1974, \aj, 79, 25
\bibitem[Ferland et al.(1998)]{fe98} Ferland, G.~J.,
Korista, K.~T., Verner, D.~A., Ferguson, J.~W., Kingdon, J.~B., \&
Verner, E.~M.\ 1998, \pasp, 110, 761
\bibitem[Finkbeiner(2003)]{fin03} Finkbeiner, D.~P.\ 2003,
\apjs, 146, 407
\bibitem[Galazutdinov et al.(2008)]{gal08} Galazutdinov,
G.~A., Lo Curto, G., \& Kre{\l}owski, J.\ 2008, \mnras, 386, 2003
\bibitem[G{\'o}rski et al.(2005)]{gor05} G{\'o}rski, K.~M.,
Hivon, E., Banday, A.~J., Wandelt, B.~D., Hansen, F.~K., Reinecke,
M., \& Bartelmann, M.\ 2005, \apj, 622, 759
\bibitem[Haffner et al.(2003)]{haf03} Haffner, L.~M.,
Reynolds, R.~J., Tufte, S.~L., Madsen, G.~J., Jaehnig, K.~P., \&
Percival, J.~W.\ 2003, \apjs, 149, 405
\bibitem[Hawley(1978)]{haw78} Hawley, S.~A.\ 1978, \apj, 224,
417
\bibitem[Herbison-Evans et al.(1971)]{her71} Herbison-Evans,
D., Hanbury Brown, R., Davis, J., \& Allen, L.~R.\ 1971, \mnras,
151, 161
\bibitem[Kalberla et al.(2005)]{ka05} Kalberla, P.~M.~W., Burton, W.~B.,
Hartmann, D., Arnal, E.~M., Bajaja, E., Morras, R., P{\"o}ppel,
W.~G.~L.\ 2005, \aap, 440, 775
\bibitem[Kilian-Montenbruck et al.(1994)]{kil94} Kilian-Montenbruck, J., Gehren, T.,
\& Nissen, P.~E.\ 1994, \aap, 291, 757
\bibitem[Korpela et al.(2006)]{kor06} Korpela, E.~J., et al.\
2006, \apjl, 644, L163
\bibitem[Kunzli et al.(1997)]{kun97} Kunzli, M., North, P., Kurucz,
R.~L., \& Nicolet, B.\ 1997, \aaps, 122, 51
\bibitem[Lasker(1977)]{las77} Lasker, B.~M.\ 1977, \apj, 212,
390
\bibitem[Mart{\'{\i}}n-Hern{\'a}ndez et al.(2002)]{mar02} Mart{\'{\i}}n-Hern{\'a}ndez, N.~L.,
et al.\ 2002, \aap, 381, 606
\bibitem[Morisset et al.(2005)]{mor05} Morisset, C.,
Stasi{\'n}ska, G., \& Pe{\~n}a, M.\ 2005, \mnras, 360, 499
\bibitem[Osterbrock(1989)]{ost89} Osterbrock, D.~E.\ 1989,
Research supported by the University of California, John Simon
Guggenheim Memorial Foundation, University of Minnesota, et al.~Mill
Valley, CA, University Science Books, 1989, 422 p.
\bibitem[Park et al.(2007)]{park07} Park, J.-W., et al.\ 2007,
\apjl, 665, L39
\bibitem[Peimbert \& Torres-Peimbert(1977)]{pei77} Peimbert, M., \&
Torres-Peimbert, S.\ 1977, \mnras, 179, 217
\bibitem[Peimbert et al.(1993)]{pei93} Peimbert, M., Storey,
P.~J., \& Torres-Peimbert, S.\ 1993, \apj, 414, 626
\bibitem[Peimbert(2003)]{pei03} Peimbert, A.\ 2003, \apj,
584, 735
\bibitem[Reynolds(1984)]{rey84} Reynolds, R.~J.\ 1984, \apj,
282, 191
\bibitem[Reynolds(1985)]{rey85} Reynolds, R.~J.\ 1985, \aj,
90, 92
\bibitem[Reynolds(1988)]{rey88} Reynolds, R.~J.\ 1988, \apj, 333,
341
\bibitem[Reynolds(2004)]{rey04} Reynolds, R. J. 2004, AdSpR, 34, 27
Tufte, S.~L., Haffner, L.~M., Jaehnig, K., \& Percival, J.~W.\ 1998,
Publications of the Astronomical Society of Australia, 15, 14
\bibitem[Reynolds et al.(2005)]{rey05} Reynolds, R.~J.,
Chaudhary, V., Madsen, G.~J., \& Haffner, L.~M.\ 2005, \aj, 129, 927
\bibitem[Rybicki \& Lightman(1979)]{ryb79} Rybicki, G.~B., \& Lightman,
A.~P.\ 1979, New York, Wiley-Interscience, 1979.~393 p.
\bibitem[Savage \& Wakker(2009)]{sav09} Savage, B.~D., \& Wakker, B.~P.\ 2009, \apj,
702, 1472
\bibitem[Sembach et al.(2000)]{sem00} Sembach, K.~R., Howk,
J.~C., Ryans, R.~S.~I., \& Keenan, F.~P.\ 2000, \apj, 528, 310
\bibitem[Stasi{\'n}ska(1982)]{sta82} Stasi{\'n}ska, G.\ 1982, \aaps, 48, 299
\bibitem[Verhamme et al.(2006)]{ver06} Verhamme, A., Schaerer, D., \&
Maselli, A.\ 2006, \aap, 460, 397
\bibitem[Wang et al.(2004)]{wan04} Wang, W., Liu, X.-W., Zhang, Y., \&
Barlow, M.~J.\ 2004, \aap, 427, 873
\bibitem[York \& Kinahan(1979)]{yor79} York, D.~G., \& Kinahan, B.~F.\ 1979, \apj,
228, 127

\end{thebibliography}
\end{document}